\documentclass[aps,pra,a4,superscriptaddress,preprintnumbers,twoside,twocolumn,floatfix]{revtex4-2}

\usepackage{import}
\usepackage{stmaryrd}
\usepackage{physics}
\usepackage{graphicx}
\usepackage{color}
\usepackage{bm,bbold}
\usepackage{bbm}
\usepackage{enumerate}
\usepackage{float}

\graphicspath{figures} 
\usepackage[caption=false]{subfig}
\usepackage{amsfonts, amsmath, amsthm, amssymb} 
\usepackage[export]{adjustbox}

\usepackage{soul}

\usepackage[colorlinks,bookmarks=false,citecolor=blue,linkcolor=red,urlcolor=blue]{hyperref}

\begin{document}

\title[ATI]{Topological quantum thermometry}

\author{Anubhav Kumar Srivastava}\email{anubhav.srivastava@icfo.eu}
\affiliation{ICFO - Institut de Ciències Fotòniques, The Barcelona Institute of Science and Technology, 08860 Castelldefels (Barcelona), Spain}

\author{Utso Bhattacharya}
\affiliation{ICFO - Institut de Ciències Fotòniques, The Barcelona Institute of Science and Technology, 08860 Castelldefels (Barcelona), Spain}

\author{Maciej Lewenstein}
\affiliation{ICFO - Institut de Ciències Fotòniques, The Barcelona Institute of Science and Technology, 08860 Castelldefels (Barcelona), Spain}
\affiliation{ICREA, Pg. Lluis Companys 23, 08010 Barcelona, Spain}

\author{Marcin Płodzień}\email{marcin.plodzien@icfo.eu}
\affiliation{ICFO - Institut de Ciències Fotòniques, The Barcelona Institute of Science and Technology, 08860 Castelldefels (Barcelona), Spain}

\begin{abstract}

An optimal local quantum thermometer is a quantum many-body system that saturates the fundamental lower bound for the thermal state temperature estimation accuracy [L. Correa, et. al., Phys. Rev. Lett. 114, 220405 (2015)]. Such a thermometer has a particular energy level structure with a single ground state and highly degenerated excited states manifold, with an energy gap proportional to the estimated temperature. In this work, we show that the optimal local quantum thermometer can be realized in an experimentally feasible system of spinless fermions confined in a one-dimensional optical lattice described by the Rice-Mele model. We characterize the system's sensitivity to temperature changes in terms of quantum Fisher information and the classical Fisher information obtained from experimentally available site occupation measurements.

\end{abstract}

\maketitle

\section{Introduction}
Quantum thermodynamics provides the necessary framework allowing building quantum thermal devices like quantum heat engines \cite{quan_2007, Kosloff_2014, Campisi_2015, Pozas-Kerstjens_2018, Peterson_2019, Myers_2022} or quantum batteries \cite{gemmer2009quantum, vinjanampathy2016quantum, goold2016role, binder2018thermodynamics, deffner2019quantum, bhattacharjee2021quantum, PhysRevResearch.2.023113, PhysRevE.105.044125,PhysRevA.101.032115,PhysRevA.104.032207,hoang2023variational,PhysRevResearch.5.013088,Koch2023}. Since most of these devices work in ultra-low temperature regimes of the order of nano- and pico-Kelvin \cite{leanhardt2003cooling, cooling_pico, horodecki2013fundamental}, there is a constant requirement for higher accuracy of the temperature estimation of such quantum systems \cite{carlos2015thermometry,PhysRevB.98.045101}.
The temperature of a given quantum system is not a quantum mechanical observable but rather a parameter of its quantum state \cite{PhysRevLett.130.040401, PhysRevB.98.104302}, and as a result, temperature estimation corresponds to the quantum estimation problem.
The accuracy of any estimation is limited by the quantum Cram\'er-Rao bound \cite{braunstein1994statistical, cramer1999mathematical} relating maximal sensitivity to the parameter changes given by quantum Fisher information, with the estimated parameter \cite{helstrom1969quantum, giovannetti2011advances, Pezz__2018, Rams2018, PhysRevA.102.012204, Demkowicz-Dobrzanski_2020,  MullerRigat2023certifyingquantum, Frerot_2023}.
The two main paradigms in quantum thermometry are based on the prior knowledge about the system's temperature. In the global quantum thermometry, there is no prior knowledge about the estimated temperature \cite{campbell2017global, mok2021optimal, rubio2021global, PRXQuantum.3.040330}. On the other hand, in the local quantum thermometry, it is assumed that the estimator for the temperature is given, whereas the aim is to minimize the uncertainty of the temperature estimator \cite{Correa2015, de2016local, de2018quantum, mehboudi2019thermometry, Nimmrichter_bayesian, mehboudi2022fundamental, Yang2024, Grattan2025}. 

The optimal local quantum thermometer, saturating the fundamental bound of the system's sensitivity to temperature changes, is a two-level many-body system with a single ground state and a degenerated manifold of the excited state, with the energy gap proportional to the estimated temperature \cite{Correa2015}.
Current theoretical efforts in local quantum thermometry are focused on finding the optimal quantum setups that maximize the sensitivity of the system's thermal state to any temperature changes. Over the recent years, different systems have been proposed as optimal quantum thermometers, including two-component fermions in a one-dimensional harmonic trap \cite{plodzien2018few, plodzien2018numerically}, thermoelectric systems \cite{thermoelectric, electronic_fluctuations}, quantum critical systems \cite{phase_transition, quantumphasetransition, criticality_thermo, anna-criticality, Aybar2022criticalquantum, critical_metrology1}, quantum dots \cite{walker2003quantum, haupt2014single, PhysRevApplied.2.024002, chekhovich2017measurement}, color centers in diamonds
\cite{Fujiwara_2021}, continuous variable systems \cite{PhysRevApplied.17.034073}, single qubit dephasing \cite{Razavian2019}, 
impurities in Bose-Einstein condensates \cite{sabin2014impurities, Correa_2017, mehboudi2019using, PhysRevResearch.4.023191,Glatthard2022bendingrulesoflow, Mihailescu2024} or Fermi gases \cite{PhysRevA.95.053627, Mitchison_2020, Oghittu_2022, Marti_latest}, as well as exotic models utilizing Unruh-DeWitt detectors \cite{Robles_2017}, utilizing the Berry phase \cite{Martin-Martinez_2013}, critical systems \cite{salvia2023critical, Ostermann2024}, and spin models \cite{ PhysRevResearch.5.043184, abiuso2023optimal, Ullah2025, Mehboudi2025, Mihailescu2023}.
In \cite{abiuso2023optimal}, with the help of machine learning techniques \cite{dawid2022modern}, the authors show that assuming only two-body interactions, the optimal local quantum thermometer can be realized with the spin-$1/2$ system in the star-spin geometry; such a system, however, is hard to realize experimentally.\\ \phantom{x}\hspace{1ex}In this work, we show that an idealized optimal local quantum thermometer with degenerated excited state, operating in the sub-nK regime, can be realized in the experimentally feasible system of $N$ spinless fermions in a one-dimensional optical lattice described by the topological Rice-Mele Hamiltonian \cite{PhysRevLett.49.1455}, with specific lattice fillings.
The proposed quantum thermometer can be tuned to the optimal temperature by varying the lattice dimerization and the staggered onsite potential. We characterize the sensitivity of the considered quantum thermometer to the temperature changes in terms of the quantum Fisher information (QFI) and the classical Fisher information (CFI) obtained from the experimentally feasible site occupation measurements. 
The Rice-Mele model in the limit of vanishing staggered potential reduces to the Su-Schrieffer-Heeger (SSH) model \cite{PhysRevLett.42.1698, RevModPhys.60.781} being one of the simplest condensed matter systems that exhibit topological characteristics (for an extensive review, see \cite{RevModPhys.83.1057, RevModPhys.82.3045, Ryu_2010, Asboth2016}). The Rice-Mele and SSH models have been experimentally realized in many quantum simulator platforms \cite{arguelloluengo2023synthetic}, such as quantum gases in optical lattices \cite{Atala2013, PhysRevB.98.245148, Lohse2015, Nakajima2016, reid2022observation, Cooper_2019, Walter2023}, acoustic systems \cite{10.1063/1.5051523}, graphene \cite{Rizzo2018}, photonic crystals \cite{PhysRevB.89.085111}, time crystals \cite{Giergiel_2019}, and in Rydberg arrays \cite{doi:10.1126/science.aav9105, Weber_2018, Lienhard:19}. There have also been recent proposals of using topological systems as potential quantum sensors \cite{McDonald2020, PhysRevResearch.4.013113, PhysRevLett.129.090503,sarkar2023quantumenhanced, Mukhopadhyay2024}.\\
\begin{figure}[t!]
\includegraphics[width=\linewidth]{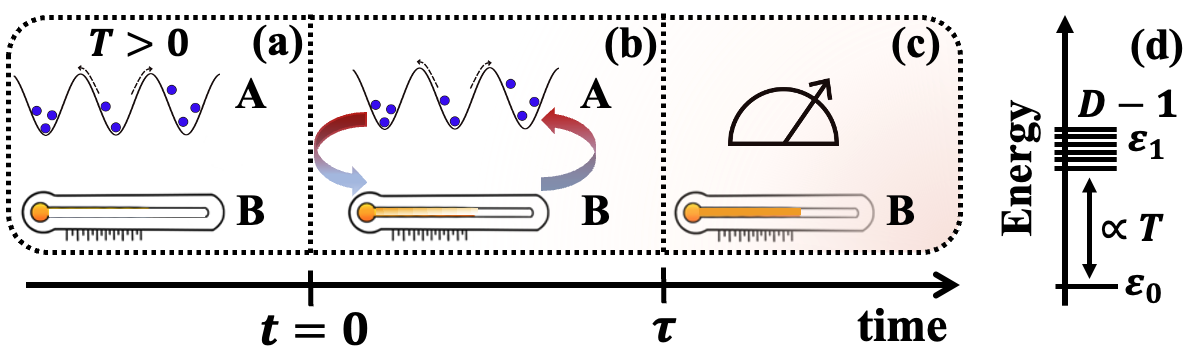}
    \caption{Schematic representation of a standard thermometry protocol. (a) The quantum thermometer is initially prepared in the pure ground state, (b) and exchanges energy with the system in the Gibbs thermal state at temperature $T$, eventually reaching a thermal state, and lastly, (c) the measurement protocol is prepared to estimate the system's temperature. Panel (d) presents schematically the energy level structure of an optimal quantum thermometer with a single ground state and $(D-1)$-fold degeneracy of the excited state of Ref.~\cite{Correa2015}.
    }
    \label{fig:Fig1}
\end{figure}

\section{Preliminaries}
Let us consider a thermal reservoir, being a quantum system described by a Hamiltonian $\hat{H}_A$ in a thermal Gibbs state $\hat{\varrho}^A_{T}$ at an unknown temperature $T$, 
and a quantum thermometer described by a Hamiltonian $\hat{H}_B$ in a pure ground state  $\hat{\varrho}^B$.
The quantum thermometer is then coupled to system A to exchange energy. After a sufficiently long time, in an ideal scenario,  the composite system reaches thermal equilibrium and is described by stationary density matrix $\hat{\varrho}_T^{AB}$. The measurement protocol is now prepared
on the reduced density matrix of the thermometer $\Tr_A[\hat{\varrho}_T^{AB}]\equiv \hat{\rho}^B_T$ 
to estimate the temperature $T$.
The full measurement protocol consists of global thermometry, i.e., estimation of $T$ without prior knowledge about temperature, and local thermometry, which minimizes the standard deviation $\Delta T$ of the estimated temperature \cite{Correa2015, mehboudi2019thermometry}, which is however lower bounded by the quantum Cram\'er-Rao bound \cite{cramer1999mathematical, giovannetti2011advances} $\Delta T \ge 1/\sqrt{{\cal N}\mathcal{F}_T}$,
where ${\cal N}$ is the prepared number of measurements and $\mathcal{F}_T$ is the quantum Fisher information  \cite{helstrom1969quantum,braunstein1994statistical, Pezz__2018}. 

The optimal local quantum thermometry aims to find a quantum system maximizing the quantum Fisher information of a thermal state with respect to changes in temperature. The quantum Fisher information for a thermal state $\hat{\rho}^B_T$ reads \cite{braunstein1994statistical, Liu_2014}
\begin{equation}
    {\cal F}_T = 4 \sum_{l,j}p_l \frac{\abs{\bra{\varepsilon_l}\partial_T\hat{\rho}^B_T\ket{\varepsilon_j}}^2}{(p_l + p_j)^2} = \frac{\Delta \hat{H}_B^2}{T^4},
    \label{eq:QFI}
\end{equation}
where $\Delta \hat{H}_B^2 \equiv \Tr[\hat{\rho}^B_T \hat{H}_B^2] - \Tr[\hat{\rho}^B_T \hat{H}_B]^2$, and $\{\varepsilon_l, |\varepsilon_l\rangle \}_{l=1}^D$
are eigenvalues and eigenvectors of $\hat{H}_B$, and $\hat{\rho}^B_T = e^{-\beta \hat{H}_B}/{\cal Z}$, where ${\cal Z} = \sum_l^D e^{-\beta\varepsilon_l}$ and $\beta = 1/k_BT$ (from here on, we introduce the natural constants $k_B=\hbar =1$), and $p_l = \bra{\varepsilon_l} \hat{\rho}^B_T\ket{\varepsilon_l} = {\cal Z}^{-1} e^{-\beta \varepsilon_l}$.

As shown in Ref.~\cite{Correa2015}, the variance $\Delta\hat{H}_B^2$ for a thermal state is maximized for the $D$-dimensional Hamiltonian $\hat{H}_B$ with a two-level energy spectrum having a single ground state, and $(D-1)$-fold degenerated excited state, with an energy gap proportional to estimated temperature, i.e.  $\varepsilon_1 - \varepsilon_0 = x T$, where $x>0$ is the solution of the equation $e^x = (D-1)(2+x)/(2-x)$.
In such a case, the QFI for an optimal local quantum thermometer \begin{equation}\label{eq:QFIoptimalextended}
      \overline{{\cal F}}_T = e^xx^2\frac{D-1}{(D-1+e^x)^2}\frac{1}{T^2} = \frac{f(D)}{T^2}.
 \end{equation} 

In the limit of large degeneracy of the excited state, i.e. $D\to \infty$, one can approximate $x\simeq \ln{D}$, $2\sqrt{f(D)}\simeq \ln{D}$, maximal QFI reads $\overline{\cal F}_T = (\ln{D}/2T)^2$,
and the relative temperature estimation accuracy is bounded from below by
$\Delta T/T \ge 2/(\sqrt{{\cal N}}\ln{D}$).

\begin{figure}[t!]
    \centering
\includegraphics[width=\linewidth]{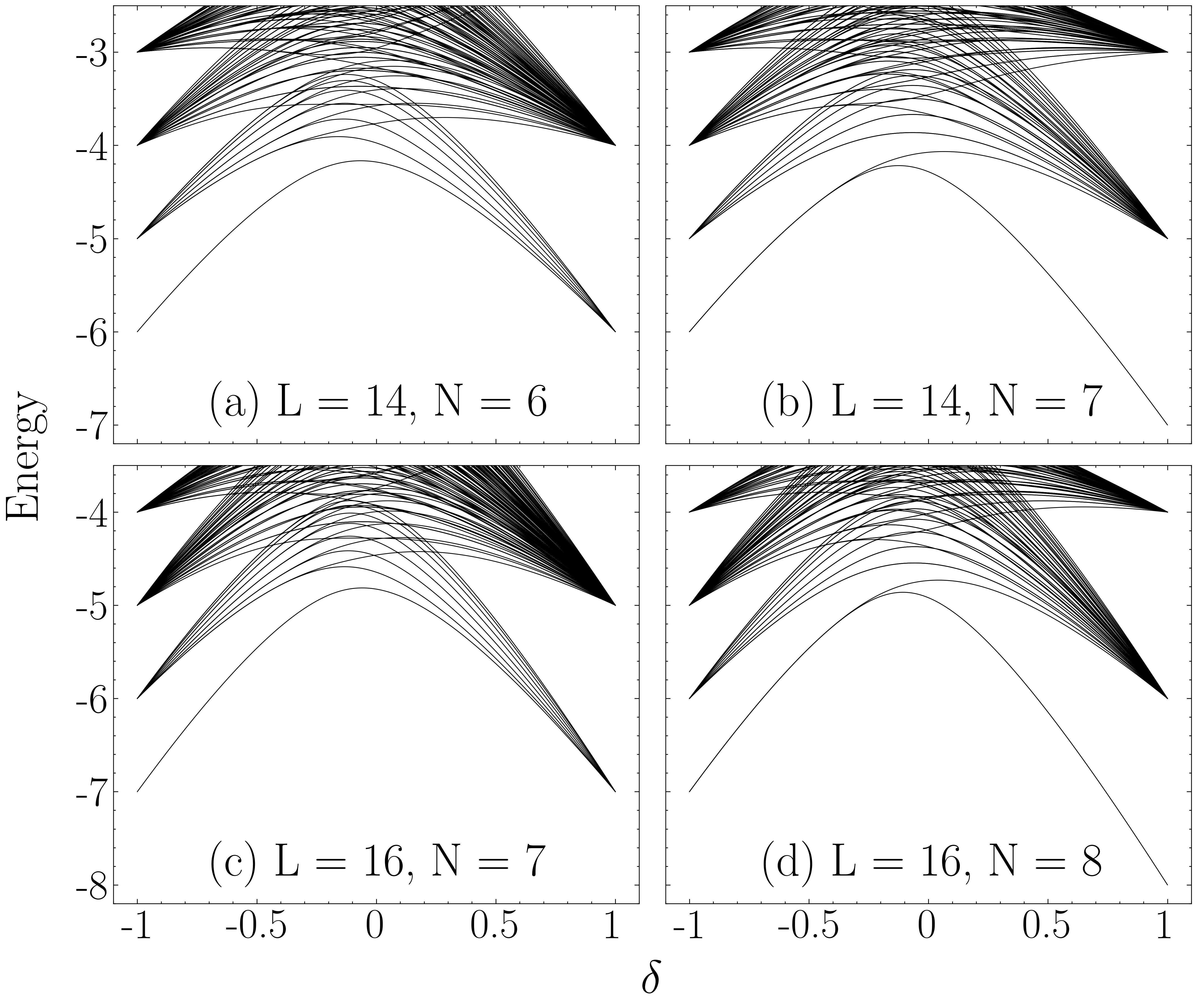}
    \caption{The energy level structure of the quantum thermometer, Eq.~\eqref{eq:HamiltonianThermometer} in the SSH limit, i.e. $m=0$, as a function of the dimerization parameter $\delta$ with $N = L/2-1$ particles (panels (a)-(c)) and with $N = L/2$ particles (panels (b)-(d)), for system sizes $L = 14$ (panels (a)-(b)) and $L = 16$ (panels (c)-(d)). 
    The optimal local quantum thermometer energy level structure is realized at dimerization limit $\delta = -1$  with $2N$-fold degeneracy in the first excited state manifold (panels (a)-(c)), and at $\delta = 1$ with $N^2$-fold degeneracy (panels (b)-(d)). 
    }
    \label{fig:Fig2}
\end{figure}

\section{Results}

\subsection{The proposed thermometer model} 

As a local quantum thermometer, we consider a system of $N$ spinless fermions in a one-dimensional lattice with $L$ sites described by the Rice-Mele Hamiltonian
\begin{equation}\label{eq:HamiltonianThermometer}
    \hat{H}_{B} = t\sum_{i}   \left(1 + (-1)^{i}\delta\right)\left(\hat{c}^{\dagger}_{i+1}\hat{c}_{i} + h.c.\right) + \frac{m}{2}\sum_{i} (-1)^i \hat{n}_i,
\end{equation}
where $\hat{c}^{\dagger}_{i} (\hat{c}_{i}$) are the fermionic creation (annihilation) operators acting on the $i$-{th} lattice site fulfilling the anti-commutation relation $\{\hat{c}_i, \hat{c}^\dagger_i \} = 1$, $\hat{n}_i = \hat{c}_i^\dagger\hat{c}_i$ is the site occupation operator, $\delta \in [-1,1]$ is the dimerization parameter, and $m\ge0$ is the staggered on-site potential strength. In the following, we set $t=1/2$.
The Hilbert space dimension is given by binomial coefficient $D = \binom{L}{N}$.
In the periodic boundary condition geometry, assuming the lattice constant of unit length, the single particle Hamiltonian, Eq.~\eqref{eq:HamiltonianThermometer}, can be expressed in the momentum representation as a two-band model $\hat{H}_B(k) = h_x(k)\hat{\sigma}_x + h_y(k)\hat{\sigma}_y + \frac{m}{2}\hat{\sigma}_z$ with $h_x(k) = t(1+\delta) + t(1-\delta)\cos k$, $h_y(k) = t(1-\delta)\sin k$, where $\hat{\sigma}_{x,y,z}$ are the Pauli operators.
At $m=0$, the Hamiltonian reduces to the SSH model and can be characterized by the topological invariant given by the winding number $\nu$ \cite{PhysRevLett.80.1800, RevModPhys.66.899, doi:10.1142/9599,Asboth2016}. The trivial dimerization limit  $\delta = 1$ corresponds to the vanishing topological invariant $\nu = 0$, while  $\delta = -1$ corresponds to the topological phase with $\nu = 1$, supporting the two zero-energy edge states in open boundary condition geometry. The experimental platform for the considered quantum system has been realized in spinful fermions using $^{171}$Yb atoms~\cite{Nakajima2016}, and more recently with potassium $^{40}$K atoms~\cite{Walter2023}. The characteristic temperature related to the recoil energy in this case is $T_0\sim200$~nK.

The Rice-Mele Hamiltonian, Eq.~\eqref{eq:HamiltonianThermometer}, can serve as a quasi-optimal local quantum thermometer with open boundary conditions for specific fillings $f=N/L$, with $N = L/2-1$ or $N = L/2$ particles, where the optimal energy gap can be tuned by the dimerization parameter $\delta$, and the staggered potential amplitude $m$.
We start with an analysis of the energy level spectrum of the system  with vanishing staggered potential amplitude $m = 0$.
We consider a lattice with $L = 14, 16$ sites and open boundary conditions with $N = L/2-1$, and $N = L/2$ particles.  With the full many-body exact diagonalization,
we calculate energy level spectra of the Hamiltonian of Eq.~\eqref{eq:HamiltonianThermometer} for each tuple $\{L, N, \delta\}$, $\delta\in [-1,1]$, represented in Fig.~\ref{fig:Fig2}. Figures \ref{fig:Fig2}(a) and \ref{fig:Fig2}(c) correspond to $N = L/2-1$ particles. The optimal energy level structure of Fig.~\ref{fig:Fig1}(d) is realized at the topological dimerization limit, $\delta = -1$, where the system has a single ground state and degenerated first excited state with $2N$-fold degeneracy. In contrast, for $\delta\ge 0$, the system enters a single quasi-degenerated manifold regime, which is also characterized by high sensitivity to temperature changes \cite{plodzien2018few}, especially at $\delta=1$ where the energy level structure resembles the one from critical thermometry Ref.~\cite{Aybar2022criticalquantum}. Figures \ref{fig:Fig2} (b) and \ref{fig:Fig2}(d) correspond to $N=L/2$ particles, where the optimal local quantum thermometer energy level structure is realized at trivial dimerization limit $\delta = 1$ with $N^2$-fold degeneracy of the first excited state. Next, we study the thermometer's sensitivity to temperature changes quantified by the quantum Fisher information.

\begin{figure}[t!]
\centering
\includegraphics[width=\linewidth]{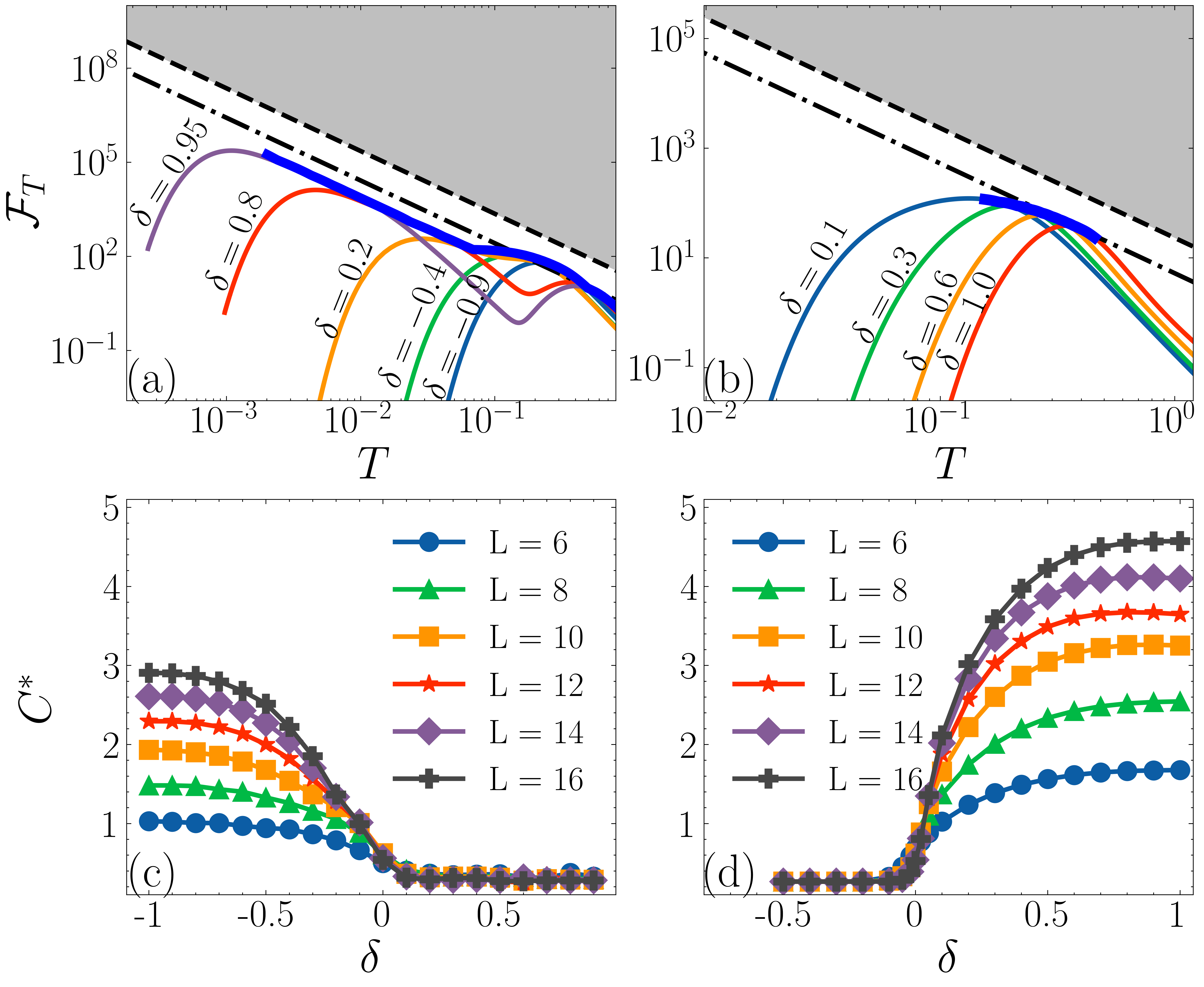}
\caption{Panels (a)-(b): (log-log scale) The sensitivity of the SSH quantum thermometer to the temperature changes in terms of QFI for $L = 16$ sites.  Solid thin lines represent ${\cal F}_T$ as a function of temperature $T$, for different values of the dimerization parameter $\delta$. The thick, solid blue line represents an envelope of the maximal QFI. The dashed line represents the QFI bound for an optimal local quantum thermometer with degeneracy corresponding to the total Hilbert space dimension $D$, while the dash-dotted line represents the QFI bound for the optimal quantum thermometer corresponding to the $2N$-fold quasi-degenerated first excited state. 
Panels (c)-(d): the maximal specific heat capacity $C^*$ (see main text) as a function of $\delta$, with increasing system size $L$. Left (right) column corresponds to $N = L/2-1$ ($N = L/2$) particles.}
\label{fig:Fig3}
\end{figure}
\subsection{Quantum Fisher information of the SSH model}

We assume that the thermometer is already in the thermal state at temperature $T$, given by $\hat{\rho}^{B}_T$.
We focus on the quantum Fisher information, ${\cal F}_T$, Eq.~\eqref{eq:QFI} to characterize the system's sensitivity to temperature changes. 

Fig.~\ref{fig:Fig3}(a)-(b) represents the quantum Fisher information ${\cal F}_T$, as a function of temperature $T$ for fixed $\delta$ (thin solid lines) for $N = L/2 - 1$ and $N = L/2$ particles respectively. The thick, solid blue line represents the maximal ${\cal F}_T$ envelope for different values of $\delta$. The thermometer's QFI is close to the QFI of the optimal local quantum thermometer 
$\overline{{\cal F}}_T$ of Eq.~\eqref{eq:QFIoptimalextended}. To compare with the optimal quantum thermometer (OQT) setup of Ref.~\cite{Correa2015}, we provide the two bounds shown by dash-dotted black line corresponding to the local two-level OQT with $2N$-fold degeneracy in the first excited manifold, while the dashed black line corresponds to the two-level OQT with $(D-1)-$ fold degeneracy in the first excited state manifold. The QFI has a peaked structure with the peak at an optimal temperature $T^* = \arg\max_T {\cal F}_T$, at which the measurement accuracy is maximized. The position of the peak of QFI can be controlled by tuning the dimerization strength $\delta$.

\begin{figure}[t!]
\centering
\includegraphics[width=\linewidth]{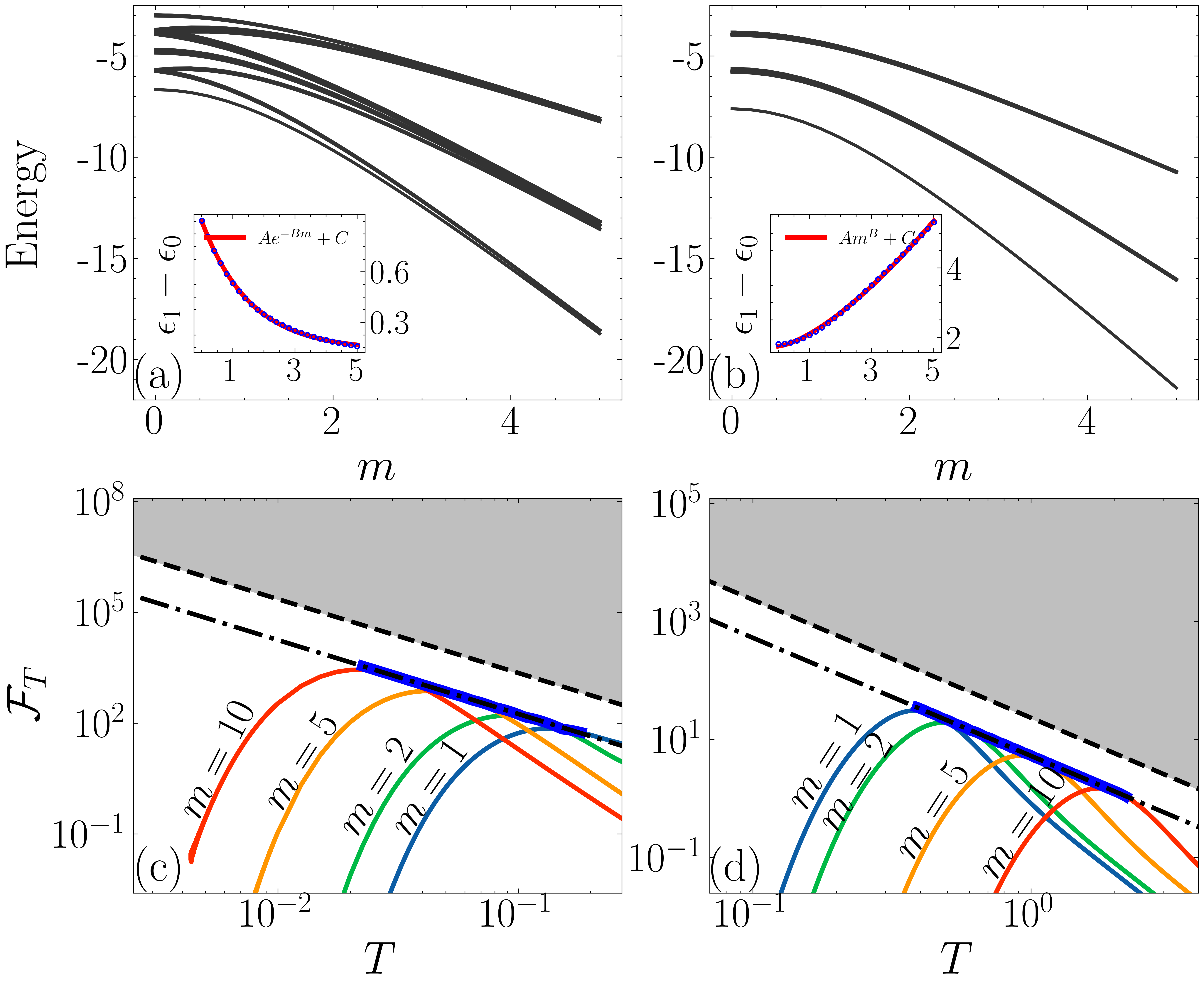}
\caption{Top row: Energy level spectrum as a function   $m$ for $L = 16$ with (a) $N = L/2-1$ and $\delta = -1$, and (b) $N= L/2$ with $\delta = 1$. Insets present the fit (solid red lines) to the energy gap (blue dots) between the ground state and the first excited energy manifold.
Bottom row: (log-log scale) quantum Fisher information ${\cal F}_T$ for various $m$ (thin solid lines) and the corresponding maximal quantum Fisher information envelope (thick solid blue line). The dashed line represents the QFI bound for an optimal local quantum thermometer \cite{Correa2015} with degeneracy in the excited state manifold corresponding to the Hilbert space dimension $D$, and the dash-dotted line denotes the QFI bound for optimal local quantum thermometer with the quasi-degeneracy corresponding to the first excited state manifold in the energy spectrum (panels (a)-(b)).
}
\label{fig:Fig4}
\end{figure}

The potential of a given thermometer for thermometry tasks is encoded into a more physically relevant quantity which is its specific heat capacity \cite{Correa2015, abiuso2023optimal}, which has been studied for phase transitions much more extensively over the recent years for spin-chain systems~\cite{Fukushima2004, mehboudi2015thermometry, Aybar2022criticalquantum, abiuso2023optimal}. Here, we consider the maximal specific heat capacity $C^*$ of the thermometer, which is defined as $C^* = {\cal F}_{T^*}T^{*2}$, where $T^*$ is the optimal temperature for a given set of model parameters $\{L,N, \delta\}$.
Fig.~\ref{fig:Fig3} (c)-(d) present $C^*$ vs $\delta$ for different system sizes $L$ with $N = L/2-1$ and $N= L/2$ particles respectively. Specific heat capacity shows the advantage of the topological dimerization regime ($\delta<0$) over the trivial one ($\delta>0$) in scaling with the system size in the $N = L/2 -1$ case, and the other way round in the $N=L/2$ case. The specific heat capacity $C^*$ scales linearly or sub-linearly with the number of spinless fermions in the chain~\footnote{$C^* \propto N^{0.81}$ for $N = L/2 - 1$ at $\delta = -1.0$, and $C^* \propto N^{1.11}$ for $N = L/2$ at $\delta = 1.0$.}, in contrast to the optimal quantum thermometer proposed in \cite{abiuso2023optimal} where scaling is quadratic (Heisenberg limit) with the number of spins, $D=2^N$. This discrepancy comes from the fact that in our protocol only fraction of the total Hilbert space forms the first excited band.

\subsection{Quantum Fisher information of the Rice-Mele model}
The crucial aspect of the practical utilization of this system as a local quantum thermometer is its energy gap tunability. Here, we show that controlling the staggered potential amplitude $m$, allows adjusting the system's sensitivity over a few orders of magnitude of temperature changes.
In Fig.~\ref{fig:Fig4} (a)-(b), we present an energy level spectrum as a function of the staggered potential amplitude $m$ with $N=L/2-1$ and $N = L/2$ particles, respectively, for $L = 16$ sites. For $N=L/2-1$ the first excited state has $N$-fold degeneracy, and $N^2$-fold degeneracy for $N=L/2$ particles. The energy gap between the ground state and the first excited state manifold decreases exponentially~\footnote{$y = Ae^{-Bm} + C$;\\ $\{A,B,C\} = \{0.76418, 0.64130, 0.13517\}$.} for $N = L/2-1$, while for $N = L/2$, the gap increases~\footnote{$y = Am^B + C$;\\ $\{A,B,C\} = \{0.35598, 1.44097, 1.74417\}$.}  with increasing staggered potential $m$.

In Fig.~\ref{fig:Fig4} (c)-(d), we present the quantum Fisher information ${\cal F}_T$ of the Hamiltonian with fixed dimerization parameter $\delta = -1$ (panel (c), $N = L/2-1$) and $\delta = 1$ (panel (d), $N = L/2$) for different values of the staggered potential amplitude $m$ (thin solid lines). The control over the staggered potential $m$  allows tuning the thermometer to the optimal temperature over three orders of magnitude by changing the position of the peak of quantum Fisher information ${\cal F}_T$.

\subsection{Measurement protocol}
Bounding the sensitivity given by the QFI requires preparing a measurement operator constructed from the eigenstates of the thermometer Hamiltonian $\hat{H}_B$, which is challenging from the experimental point of view. 
To infer the sensitivity of the thermal state $\hat{\rho}^B_T$ of the thermometer to temperature changes from the measurements easily accessible in the experiment, we focus on the classical Fisher information  $F_T = \sum_i \frac{1}{p_i(T)}\left( \frac{dp_i(T)}{dT}\right)^2$,
where $p_i(T)$ are the probabilities of measurement outcomes. Here, as a measurement protocol, we consider the normalized lattice site occupation $p_i(T) = \Tr[\hat{n}_i\hat{\rho}^B_T]/N$.

\begin{figure}[t!]
\centering
\includegraphics[width =\linewidth]{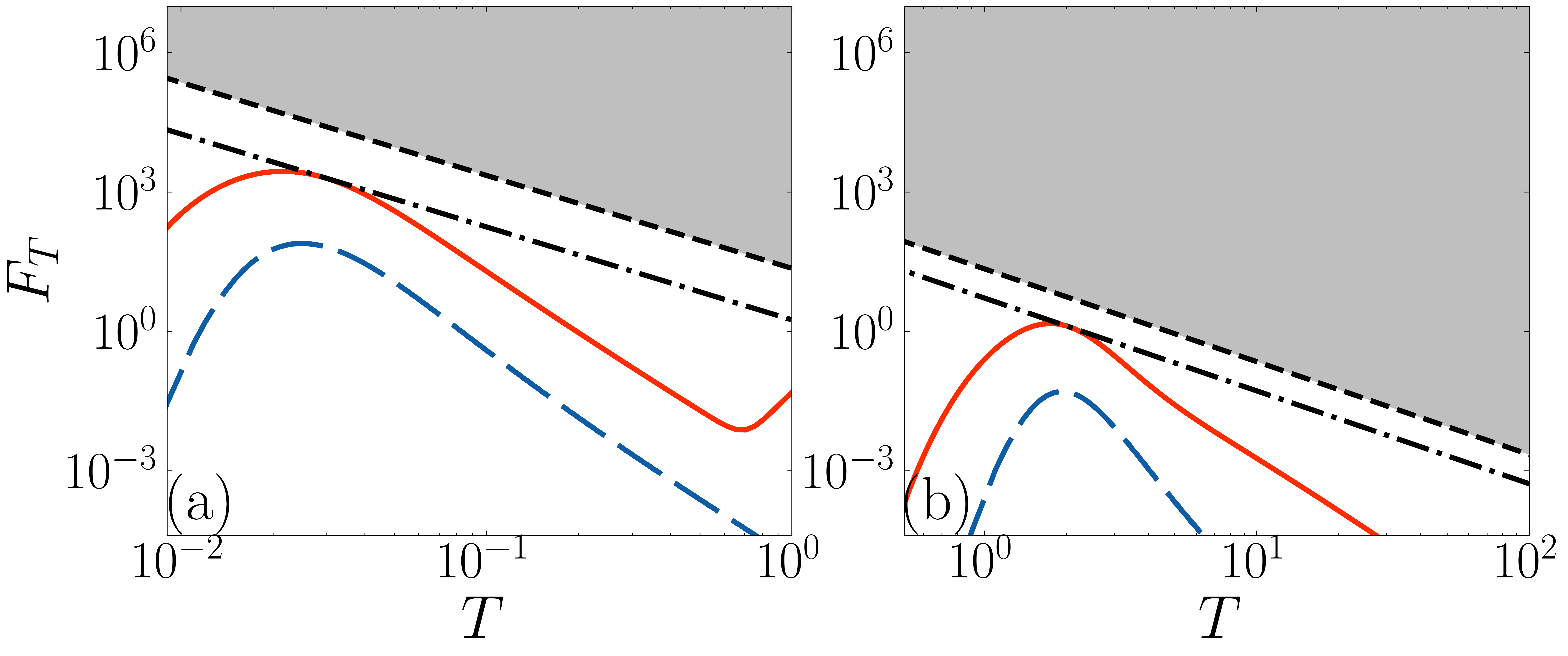}
\caption{Classical and quantum Fisher information for $L = 16$ lattice sites with (a) $N = L/2-1$ particles, at the topological dimerization limit $\delta = -1$, with  $m = 10$, and (b) $N = L/2$ particles, at the trivial dimerization limit $\delta = 1$, and   $m = 10$. The dashed blue lines represent the classical Fisher information, $F_T$, obtained from lattice site occupation measurements, while the red solid line represents the quantum Fisher information, ${\cal F}_T$, of the system. The dashed and dash-dotted lines denote the same as in Fig.~\ref{fig:Fig4}.} \label{fig:Fig5}
\end{figure}

In Fig.~\ref{fig:Fig5}, we show the classical Fisher information $F_T$ for $L = 16$ sites, with panel (a) $N=L/2-1$ particles at topological dimerization limit $\delta=-1$ with  $m=10$, and panel (b) $N = L/2$ at $\delta=1$ with $m = 10$.  
The peak in the classical Fisher information $F_{T^*}$ obtained from the experimentally feasible site occupation measurements is of the order of $\sim 10^2$, indicating the high sensitivity of the site occupation measurements to temperature changes, highlighting the usefulness of the proposed system as a quantum thermometer. 

\begin{figure}[t!]
\centering
\includegraphics[width = \linewidth]{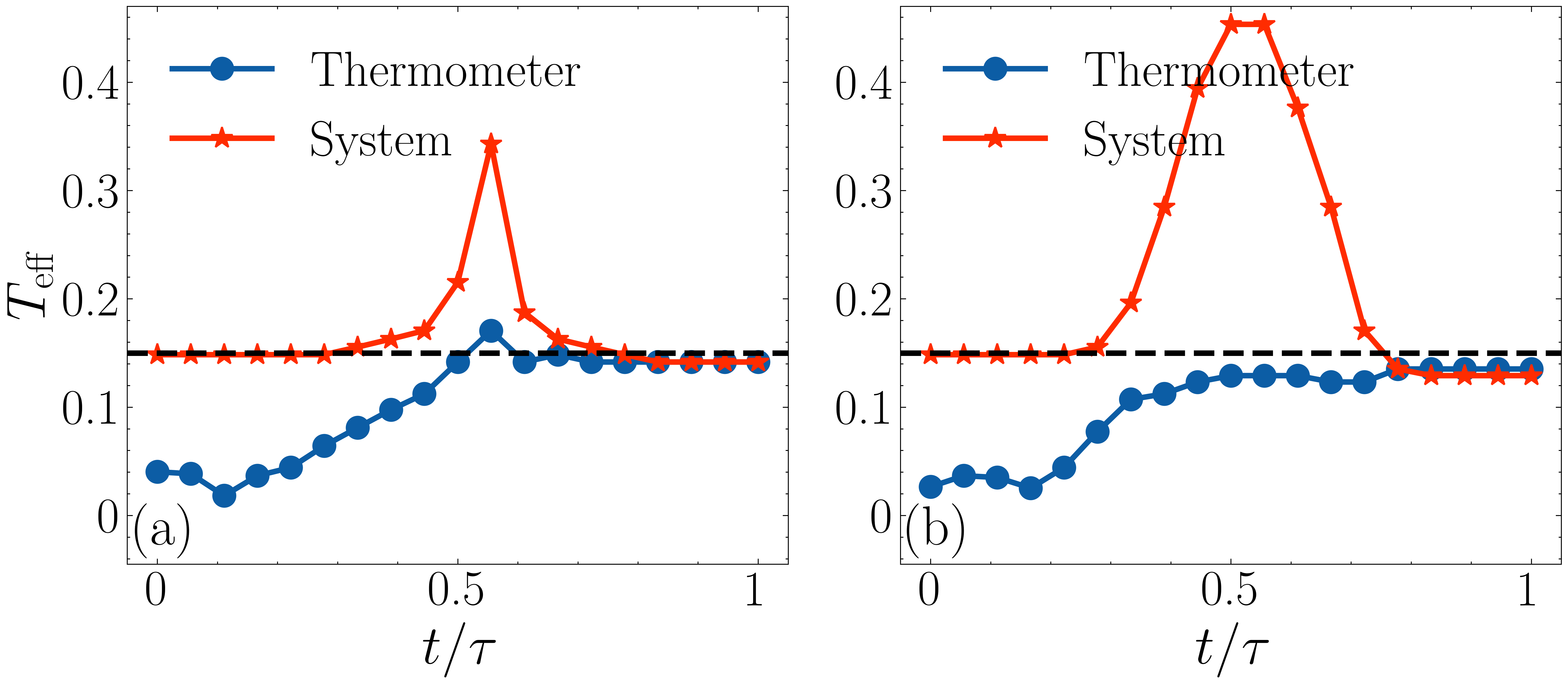}
\caption{We show the effective temperature during the time evolution of the SSH fermionic thermometer probe $(N=3, L=8)$, in the topological phase $(\delta =-0.9)$, and (a) SSH fermionic system $(N=7, L=8)$ and (b) bosonic system $(N=2,L=8)$ given by the Hamiltonian from Eq.~\eqref{eq:Hamiltonianbosons} for optimum interaction strengths $V_{AB}$ and the duration of interaction.}
\label{fig:Fig8}
\end{figure}

\section{Thermometer thermalization}

In the following, we study the dynamics of the thermometer coupled to a system, towards a thermalized state.
We prepare initial state of the total system in the product state $\hat{\rho}^{AB}(t=0^{-}) = |GS\rangle\langle GS|\otimes\hat{\rho}^{B}_T$, where thermometer is in a pure ground state $|GS\rangle$, while the system is in a Gibbs state at $T>0$. Next, we couple the thermometer to the system via on-site contact interactions,
\begin{equation}
\hat{H}_{AB}(t) = V_{AB}(t)\sum_{i=1}^{L}\hat{n}_{i,A}\hat{n}_{i,B},
\end{equation}
where $\hat{n}_{i,A/B}$ are the number operators acting on the $i-$th site of the subsystem$-A$ or $B$.
The time dependent coupling $V_{AB}(t)$ is a slowly changing function over time interval $\tau$ given by,
\begin{equation}
V_{AB}(t)= V_0  \frac{2}{\tau}\left(|t-\tau/2| -\tau/2\right)\ \forall\ t \in [0,\tau].
\end{equation}

Next, we numerically solve
\begin{equation}
\begin{split}
\dot{\hat{\rho}}^{AB}(t) & = -i[\hat{H}(t), \hat{\rho}^{AB}(t)],\\
 \hat{H}(t) &= \hat{H}_A \otimes \mathbb{I}_B+ \mathbb{I}_A \otimes \hat{H}_B + \hat{H}_{AB}(t),
 \end{split}
\end{equation}
with initial condition given by $\hat{\rho}^{AB}(t=0^-)$, and calculate reduced density matrices for thermometer and the system $\hat{\rho}^{A,B}(t) = \Tr_{A,B}[\hat{\rho}^{AB}(t)]$. Finally, we define the temperature of subsystems $A,B$ as
\begin{equation}
T^{A,B}_{\rm eff}(t) \equiv \text{arg max}_{T} \mathbb{F}(\hat{\rho}^{A,B}(t), \hat{\varrho}^{A,B}_{T}),
\end{equation}
where $\mathbb{F}(\cdot,\cdot)$ is a fidelity between two density matrices, and $\hat{\varrho}^{A,B}_T$ is a Gibbs state for Hamiltonian $\hat{H}_{A,B}$.

Fig.~\ref{fig:Fig8} illustrates the thermalization dynamics, i.e. time evolution of the effective temperature $T_{\rm eff}$ of the SSH thermometer probe when coupled to two distinct systems: (i) another SSH chain and (ii) a Bose-Hubbard chain, characterized by the Hamiltonian:
\begin{equation}\label{eq:Hamiltonianbosons}
    \hat{H}_{BH} = -J \sum_{\langle i,j\rangle} (a_{i}^{\dagger}a_j + a_j^{\dagger}a_i) + \frac{U}{2} \sum_i \hat{n}_i\hat{n}_{i+1},
\end{equation}
where $a_i^{\dagger}(a_i)$ are the bosonic creation and annihilation operators, and $\hat{n}_i$ is the number operator on the $i-$th site. By appropriately tuning the interaction strength $V_{AB}$ and the duration $\tau$, we observe that the thermometer probe equilibrates to a temperature closely matching the temperature of the system without affecting the initial temperature of the system. The near-optimal case, wherein the probe accurately reflects the system's temperature, is presented in the figure.

\begin{figure}[t!]
\centering
\includegraphics[width =\linewidth]{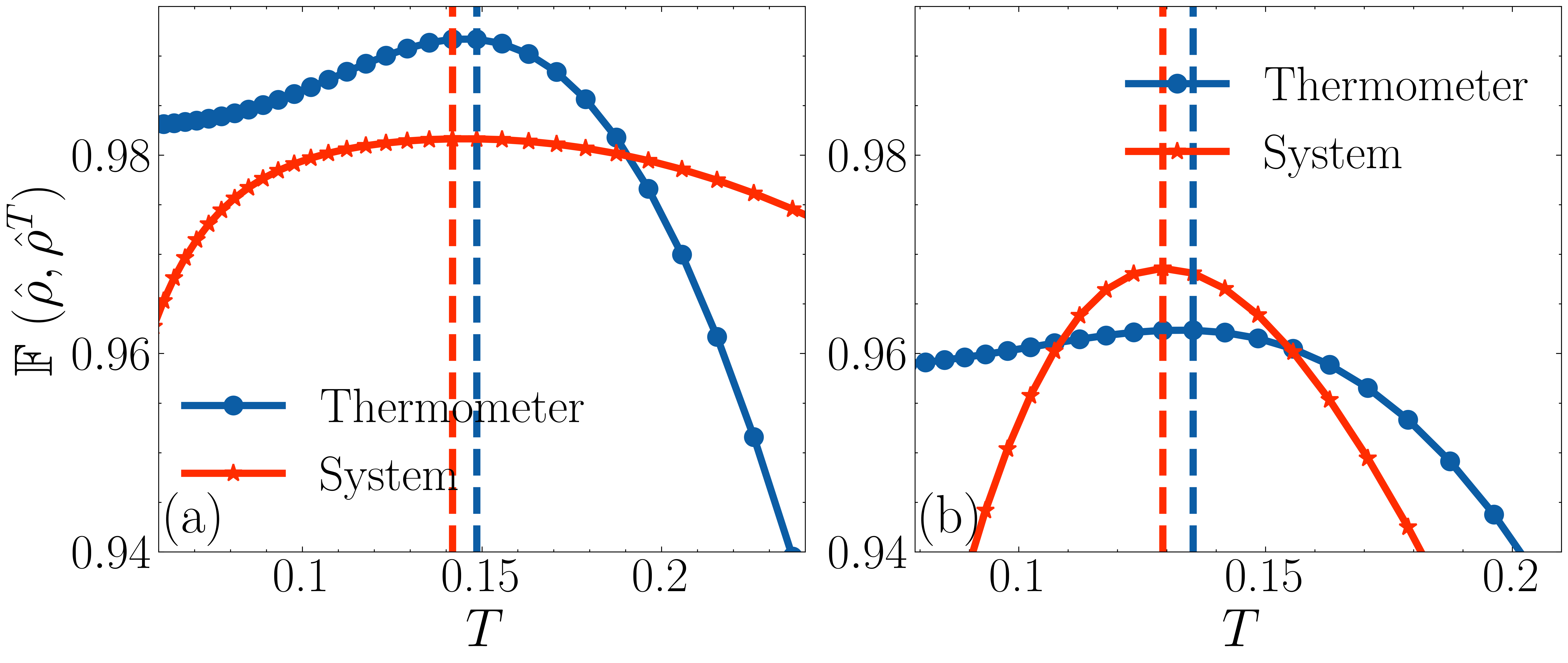}
\caption{The fidelity of the time evolved reduced density matrices with respect to the thermal states at different temperatures of the SSH fermionic thermometer probe and (a) another SSH fermionic system, and the (b) bosonic Bose-Hubbard system of Eq.~\eqref{eq:Hamiltonianbosons}. The maxima in the fidelity gives us the thermal state which is closest to the subsystem, and is used to estimate the temperature.} \label{fig:FigApp1}
\end{figure}

Fig.~\ref{fig:FigApp1} presents the fidelity of reduced density matrices $\hat{\rho}^{A,B}(t=\tau)$ with the Gibbs states for the subsystems $A,B$, at the end of time evolution, $t=\tau$. The fidelity between the reduced states and the thermal states maximizes for a particular temperature $T$ of a thermal state of $\hat{\varrho}^{A,B}_T$, denoted by the vertical dashed lines. The maxima in the fidelity between the states is $>0.96$, indicating that reduced density matrices are of Gibbs-type.

\begin{figure}[t!]
\centering
\includegraphics[width =\linewidth]{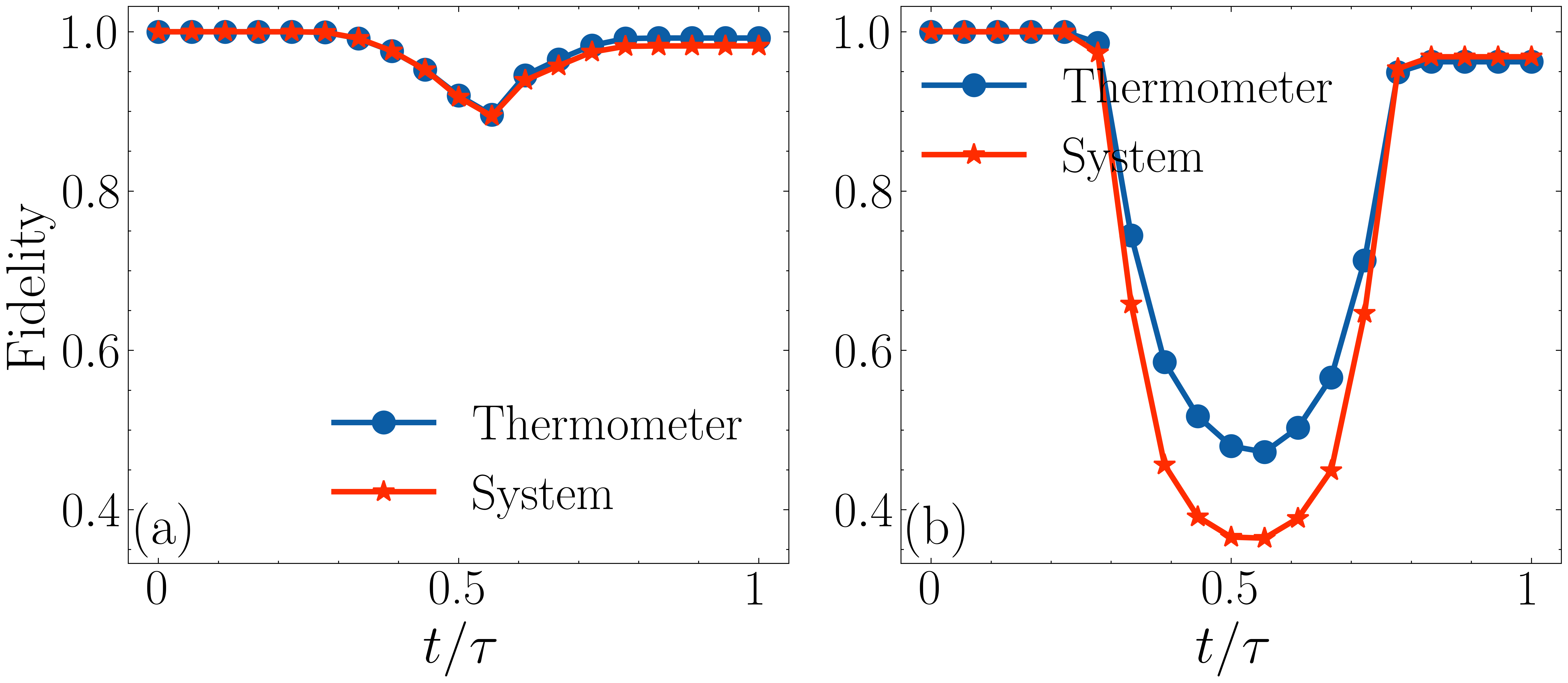}
\caption{Maximum fidelity vs time for the reduced subsystems for the (a) fermionic SSH chain system, and (b) bosonic Bose-Hubbard system of Eq.~\eqref{eq:Hamiltonianbosons} with respect to the closest thermal state during the time evolution. We see that the initial fidelity when there are no interactions is close to $1$ which is expected as the states are prepared in thermal states. The switching on of the interactions leads to a fall in the fidelity, but when the interactions are switched off the maximum fidelity is $>0.96$, validating the operational definition of temperature that we have considered.} \label{fig:FigApp2}
\end{figure}

To further understand the dynamics during the time evolution, we show in Fig.~\ref{fig:FigApp2}, the evolution of maximum fidelity with time.
The maximum fidelity between the reduced states and the set of thermal states is $1$ which is expected as the states are prepared in thermal states initially, and during the evolution the maximum fidelity falls depending on the strength of the time dependent interaction and finally, after the interactions are switched off, we see that there exists a thermal state close to the reduced states.

\section{Conclusions}
We showed that the experimentally feasible system of spinless fermions confined in a one-dimensional optical lattice described by the Rice-Mele model can serve as an optimal local quantum thermometer operating in the sub-nK temperature regime. 
Moreover, we showed that the classical Fisher information obtained from  experimentally feasible lattice site occupation measurements is close to the quantum Fisher information limit. Finally, the utility of the proposed thermometer probe is demonstrated through an investigation of its thermalization dynamics with a coupled system. The results indicate that the probe equilibrates to the system's temperature by appropriately tuning the interaction strength and the duration of interaction without affecting the temperature of the system. This highlights the efficacy of the proposed model as a reliable equilibrium quantum thermometer.

{\it Acknowledgements}---
We thank Grzegorz Rajchel-Mieldzioć and Arkadiusz Kosior for carefully reading the manuscript. We also thank Guillem Müller-Rigat and Mohammad Mehboudi for the insightful discussions and comments. 
A.K.S. acknowledges support from the European Union’s Horizon 2020 Research and Innovation Programme under the Marie Skłodowska-Curie Grant Agreement No. 847517.
M.P. acknowledges the support
of the Polish National Agency for Academic Exchange, the Bekker programme no: PPN/BEK/2020/1/00317.
The ICFO group acknowledges support from ERC AdG NOQIA; MCIN/AEI (PGC2018-0910.13039/501100011033,  CEX2019-000910-S/10.13039/501100011033, Plan National FIDEUA PID2019-106901GB-I00, Plan National STAMEENA PID2022-139099NB-I00, project funded by MCIN/AEI/10.13039/501100011033 and by the “European Union NextGenerationEU/PRTR” (PRTR-C17.I1), FPI); QUANTERA DYNAMITE PCI2022-132919, QuantERA II Programme co-funded by European Union’s Horizon 2020 program under Grant Agreement No 101017733; Ministry for Digital Transformation and of Civil Service of the Spanish Government through the QUANTUM ENIA project call - Quantum Spain project, and by the European Union through the Recovery, Transformation and Resilience Plan - NextGenerationEU within the framework of the Digital Spain 2026 Agenda; Fundació Cellex; Fundació Mir-Puig; Generalitat de Catalunya (European Social Fund FEDER and CERCA program; Barcelona Supercomputing Center MareNostrum (FI-2023-3-0024); HORIZON-CL4-2022-QUANTUM-02-SGA;  PASQuanS2.1, 101113690, EU Horizon 2020 FET-OPEN OPTOlogic, Grant No 899794, QU-ATTO, 101168628;  EU Horizon Europe research and innovation program under grant agreement No 101080086 NeQSTGrant Agreement 101080086 — NeQST); ICFO Internal “QuantumGaudi” project. Views and opinions expressed are however those of the author(s) only and do not necessarily reflect those of the European Union, European Commission, European Climate, Infrastructure and Environment Executive Agency (CINEA), or any other granting authority. Neither the European Union nor any granting authority can be held responsible for them.

\bibliography{biblio}
\end{document}